\documentclass[a4paper,10pt]{article}       
\usepackage[utf8]{inputenc}
\usepackage{amsmath}
\pdfoutput=1
\usepackage{amssymb}
\usepackage{array}
\usepackage{bbold}
\usepackage{csquotes}
\usepackage{tcolorbox}
\usepackage{graphicx}
\usepackage{pdfpages}
\usepackage[square,numbers]{natbib}
\usepackage{url}
\usepackage[top=1in, bottom=1.25in, left=1.25in, right=1.25in]{geometry}
\numberwithin{equation}{section}
\usepackage{chngcntr}
\usepackage{booktabs}
\usepackage{mathtools}
\counterwithin{table}{section}
\usepackage{blindtext}

\begin{document}

    \begin{center}
        \vspace*{0.2cm}
        
        \large
        \textbf{Deep learning facilitates fully automated brain image registration of optoacoustic tomography and magnetic resonance imaging}
        
        \vspace{1.5cm}
        
        \normalsize

		\centerline{\scshape Yexing Hu}
		\medskip
		{\footnotesize
		 \centerline{School of Information Science and Technology, ShanghaiTech University}
		   \centerline{Shanghai, China}
		} 

		\medskip

		\centerline{\scshape Berkan Lafci}
		\medskip
		{\footnotesize
		 \centerline{Institute for Biomedical Engineering and Institute of Pharmacology and Toxicology, Faculty of Medicine, University of Z\"urich}
		   \centerline{Z\"urich, Switzerland}
		 \centerline{Institute for Biomedical Engineering, Department of Information Technology and Electrical Engineering, ETH Z\"urich and University of Z\"urich}
		   \centerline{Z\"urich, Switzerland}
		}

		\medskip

		\centerline{\scshape Artur Luzgin}
		\medskip
		{\footnotesize
		 \centerline{Institute for Biomedical Engineering and Institute of Pharmacology and Toxicology, Faculty of Medicine, University of Z\"urich}
		   \centerline{Z\"urich, Switzerland}
		 \centerline{Institute for Biomedical Engineering, Department of Information Technology and Electrical Engineering, ETH Z\"urich and University of Z\"urich}
		   \centerline{Z\"urich, Switzerland}
		}
		
		\medskip

		\centerline{\scshape Hao Wang}
		\medskip
		{\footnotesize
		 \centerline{Institute for Biomedical Engineering and Institute of Pharmacology and Toxicology, Faculty of Medicine, University of Z\"urich}
		   \centerline{Z\"urich, Switzerland}
		 \centerline{Institute for Biomedical Engineering, Department of Information Technology and Electrical Engineering, ETH Z\"urich and University of Z\"urich}
		   \centerline{Z\"urich, Switzerland}
		}
		
		\medskip

		\centerline{\scshape Jan Klohs}
		\medskip
		{\footnotesize
		 \centerline{Institute for Biomedical Engineering, Department of Information Technology and Electrical Engineering, ETH Z\"urich and University of Z\"urich}
		   \centerline{Z\"urich, Switzerland}
		}
		
		\medskip
		
		\centerline{\scshape Xosé Luís Deán-Ben}
		\medskip
		{\footnotesize
		 \centerline{Institute for Biomedical Engineering and Institute of Pharmacology and Toxicology, Faculty of Medicine, University of Z\"urich}
		   \centerline{Z\"urich, Switzerland}
		 \centerline{Institute for Biomedical Engineering, Department of Information Technology and Electrical Engineering, ETH Z\"urich and University of Z\"urich}
		   \centerline{Z\"urich, Switzerland}
		}
		
		\medskip
		
		\centerline{\scshape Ruiqing Ni*}
		\medskip
		{\footnotesize
		 \centerline{Institute for Regenerative Medicine, University of Zurich}
		   \centerline{Z\"urich, Switzerland}
		 \centerline{Institute for Biomedical Engineering, ETH Zurich and University of Zurich, Zurich, Switzerland}
		   \centerline{Z\"urich, Switzerland}
		}
		
		\medskip
		
		\centerline{\scshape Daniel Razansky*}
		\medskip
		{\footnotesize
		 \centerline{Institute for Biomedical Engineering and Institute of Pharmacology and Toxicology, Faculty of Medicine, University of Z\"urich}
		   \centerline{Z\"urich, Switzerland}
		 \centerline{Institute for Biomedical Engineering, Department of Information Technology and Electrical Engineering, ETH Z\"urich and University of Z\"urich}
		   \centerline{Z\"urich, Switzerland}
		}
		
		\medskip
		
		\centerline{\scshape Wuwei Ren*}
		\medskip
		{\footnotesize
		 \centerline{School of Information Science and Technology, ShanghaiTech University}
		   \centerline{Shanghai, China}
		}

		\bigskip


        \vspace{1cm}

		\footnotetext[2]{Yexing Hu and Berkan Lafci contributed equally to this work.} 
		\footnotetext[3]{*Corresponding author: Ruiqing Ni (email:ruiqing.ni@uzh.ch), Daniel Razansky (e-mail:daniel.razansky@uzh.ch) and Wuwei Ren (e-mail:renww@shanghaitech.edu.cn).}
		\footnotetext[4]{Ruiqing Ni received funding from Synapsis foundation career development award (2017 CDA-03), Vontobel Stiftung, Helmut Horten Stiftung and UZH Entrepreneur Fellowship of the University of Zurich, reference no. [MEDEF-20 021]. Daniel Razansky received funding from Swiss Data Science Center (C19-04). Wuwei Ren received funding from ShanghaiTech University.  }
		
    \end{center}
    \begin{abstract} 
\vspace{0.5cm}
Multi-spectral optoacoustic tomography (MSOT) is an emerging optical imaging method providing multiplex molecular and functional information from the rodent brain. It can be greatly augmented by magnetic resonance imaging (MRI) that offers excellent soft-tissue contrast and high-resolution brain anatomy. Nevertheless, registration of multi-modal images remains challenging, chiefly due to the entirely different image contrast rendered by these modalities. Previously reported registration algorithms mostly relied on manual user-dependent brain segmentation, which compromised data interpretation and accurate quantification. Here we propose a fully automated registration method for MSOT-MRI multimodal imaging empowered by deep learning. The automated workflow includes neural network-based image segmentation to generate suitable masks, which are subsequently registered using an additional neural network. Performance of the algorithm is showcased with datasets acquired by cross-sectional MSOT and high-field MRI preclinical scanners. The automated registration method is further validated with manual and half-automated registration, demonstrating its robustness and accuracy.
\end{abstract}


\section{Introduction}

Multispectral optoacoustic tomography (MSOT) has provided new functional and molecular imaging capabilities by combining the rich contrast of optical imaging with the high resolution of ultrasound at depths up to several centimeters within biological tissues \cite{b1}. This has been exploited in preclinical studies involving the use of small animals (mice) e.g. to visualize vascular anatomy and hemodynamic responses, cancer angiogenesis or neuronal activity \cite{b2,b3,b4,b5,b6,b7,b8,b9}. MSOT was further shown to be a valuable tool in several clinical applications such as the diagnosis of breast cancer or melanomas \cite{b10,b11,b12}. The operational principle of MSOT is based on tissue excitation with short laser pulses at multiple optical wavelengths. This enables differentiating spectrally-distinctive endogenous and extrinsically-administered contrast agents via spectral unmixing, which represents a powerful approach for visualizing molecular activity \cite{b13,b14,b15}. However, despite its wide use in biomedical research applications \cite{b16}, \emph{in vivo}  MSOT suffers from poor soft-tissue contrast, and identification of organs or tissue types remains challenging. In comparison, magnetic resonance(MR) imaging (MRI) is a well-established high-resolution imaging modality arguably providing the richest soft-tissue contrast \cite{b17}. MRI is capable of delivering multiplex structural information exploiting different mechanisms of contrast generation \cite{b18}, which can help localizing the specific molecular information derived from MSOT signals. A multimodal strategy combining MSOT and MRI can then offer highly complementary information, and an accurate and robust registration algorithm capable of matching these two modalities can be of great value in many biomedical fields.

\begin{figure*}[!t]
\centerline{\includegraphics[width=\textwidth]{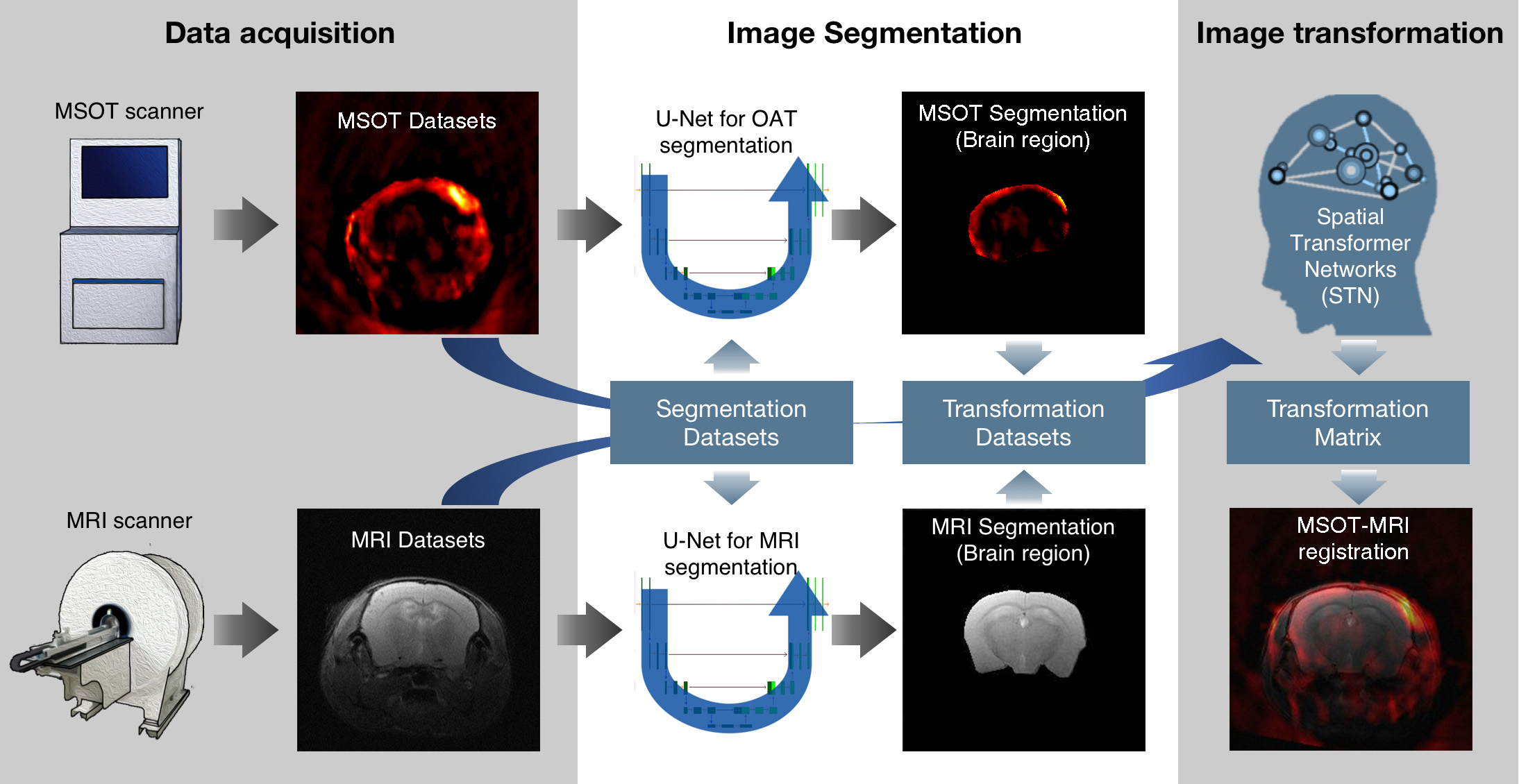}}
\caption{Figure 1: Workflow for MSOT-MRI registration. Three steps of the registration process, namely data acquisition, image segmentation and transformation are shown. Data acquisition: multi-spectral optoacoustic tomography (MSOT) and magnetic resonance imaging (MRI) datasets were acquired with a MSOT system equipped with a ring-shaped or arc-shaped ultrasound transducer and a high-field preclinical MRI scanner respectively (left panel). The training datasets for follow-up segmentation and transformation were reconstructed using the acquired raw signals. Image segmentation: the brain region of a mouse was segmented by using a U-Net-like architecture with a selected convolution kernel and padding size for different input and output on each dataset (middle panel). Image transformation: the segmented MSOT and MR images were used for training a spatial transformer networks (STN)-like network and yield a transformation matrix that can map an MSOT image onto the MRI reference. An overlaid MSOT-MR image rendered with the deep-learning(DL)-based registration method is shown (right panel).}
\label{fig1}
\end{figure*}

Previously-reported MSOT-MRI registration was based on manual registration to certain extent \cite{b19,b20,b21}. A straightforward piecewise linear mapping algorithm to co-register MSOT and MRI images was proposed to solve the registration problem \cite{b22}. The concentration maps of different fluorescence probes and hemoglobin provided by MSOT can be localized with the anatomical reference given by MRI. The transformation matrix was determined based on manually selected fiducial markers.We recently developed a semi-automated MSOT-MRI registration toolbox named RegOA based on self-adaptive mutual information (MI) \cite{b23}. The accuracy and robustness of RegOA has been demonstrated using \emph{in vivo} and \emph{ex vivo} mouse brain data. Alternatively, an integrated framework for registration of MSOT and MRI data based on combining a novel MRI animal holder and a landmark-based software co-registration algorithm was also proposed \cite{b24}. The protocol was shown to significantly improve the registration between the two modalities for both the entire body and a localized tumor, thus enabling more precise multi-modal tumor characterization and staging. Nevertheless, the first step of the RegOA protocol requires preprocessing and segmentation of MSOT \cite{b25} and MRI images, for which an initial rough estimation of a brain contour is defined by the users. Subsequently, an iterative optimization method refines the segmentation results after the initial guess. Thus, the MSOT-MRI registration methods mentioned above involve human intervention to a certain degree, hence the registration accuracy and speed are highly dependent on the experience of the users.

Herein, we introduce a deep-learning(DL)-based method for the registration of MSOT and MR images aiming to reduce manual intervention to a minimum and improve the registration accuracy. To the best of our knowledge, a similar approach has never been suggested. Unlike the previously-reported registration techniques mentioned above, a properly trained deep-neural-network-based algorithm needs no prior assumption or user experience. Thereby, the proposed method can be fully automated and greatly enhance the throughput capacity of multi-modal MSOT-MR imaging.
 
\section{Methods}
The MSOT-MRI registration procedure consists of three steps (Fig. 1), i.e., 1) acquisition of MSOT and MRI images and preparation of training datasets, 2) segmentation of the brain region on the MSOT and MRI datasets, and 3) image transformation mapping MSOT data onto MRI data. Separate deep neural networks were developed for steps 2 and 3 as described below. The registration results were evaluated by comparing the positions of manually selected reference landmarks in both images.

\subsection{Data acquisition and preparation of MSOT and MR images}
C57BL/6 mice (N=18, aged 16 month-old, both gender) were used to acquire MSOT and MR images as described previously \cite{b26}. Animals were housed in ventilated cages inside a temperature-controlled room under a 12-h dark/light cycle. Pelleted food (3437PXL15, CARGILL) and water were provided \emph{ad-libitum}. All experiments were performed in accordance with the Swiss Federal Act on Animal Protection and were approved by the Cantonal Veterinary Office Zurich (permit number: ZH090-16). All procedures fulfilled the ARRIVE 2.0 guidelines on reporting animal experiments \cite{b27}.

All MSOT images were acquired by using a commercial MSOT system (inVision 128, iThera Medical, Germany) described in detail elsewhere \cite{b19,b28,b29}. Briefly, a tunable pulsed laser was used to illuminate the object with wavelengths between 680 nm and 980 nm. The generated ultrasound waves were detected by 128 cylindrically distributed transducers with a central frequency of 5 MHz and bandwidth of 60 \%. The raw data was reconstructed by using a filtered back-projection algorithm implemented on a graphics processing unit (GPU). Imaging was performed at five wavelengths (715, 730, 760, 800 and 850 nm) on coronal slices, with 10 averages per slice. All images performed at these five wavelengths were used for deep-learning-based method. Rostral caudal brain coverage was extended over a length of approximately 12 mm with a step size of 0.3 mm resulting 40 slices. The reconstructed MSOT images were analyzed using single wavelength without any spectral unmixing.

MRI data acquisition was performed on a 7/16 small animal MRI scanner (Bruker Biospin GmbH, Ettlingen, Germany) equipped with an actively shielded gradient set of 760 mT/m with 80 ms rise time and operated by a Paravision 6.0 software platform (Bruker Biospin GmbH, Germany) \cite{b30,b31,b32}. A circular polarized volume resonator for signal transmission and an actively decoupled mouse brain quadrature surface coil with integrated combiner and preamplifier for signal receiving were used. T$_{2}$-weighted anatomical reference images were acquired in coronal and sagittal orientations. A spin-echo sequence was used with rapid acquisition relaxation enhancement, echo time = 33 ms; relaxation time = 2500 ms; rapid acquisition relaxation enhancement factor = 8; 15 sagittal slices of 1 mm thickness; interslice distance = 0, field-of-view (FOV) = 20 × 20 mm; image matrix = 256 × 256; spatial resolution = 78 × 78 mm. For each mouse brain in the current study, a T$_{2}$-weighted anatomical image of 1 mm thickness was acquired approximately at Bregma 
- 1.46 mm with the current setting of FOV.

The datasets used in this study consist of three parts: (1) T$_{2}$-weighted anatomical MR images of mouse brains and the corresponding segmented brain masks; (2) Anatomical MSOT images of mouse brains and the corresponding segmented brain masks; (3) Ground truth of registered MSOT-MR image pairs. For Dataset-1, 69 slices of T$_{2}$-weighted MR images and the corresponding brain masks were selected from 18 mice. The segmentation was performed by applying an active contour model \cite{b33}. To increase training dataset variability, data augmentation methods were used on MR images by applying rotation and scaling operations \cite{b34}, which yielded 552 pairs of MR images and the corresponding masks. For Dataset-2, 415 anatomy-segmentation pairs were generated out of the raw MSOT images acquired from 5 mice. For Dataset-3, the ground truth of the registration was generated by using a MI-based method \cite{b23}. To ease the data processing, the image dimensions of all MSOT and MR images, including the masks, were set to 256 × 256. 

\subsection{Deep-learning-based segmentation of the brain}
Three networks were involved in the deep-learning-based MSOT-MRI registration pipeline (Fig. 1). We used PyTorch 3.8.0 with CUDA 11.0 on desktop computer with the CPU i9 - 10980XE (3 GHz) and GPU RTX 3090 (10496 cores) to build and train all networks.

The MRI segmentation network is schematically depicted in Fig. 2. The network was based on the U-Net structure, which was initially proposed to solve segmentation problems appeared with a limited number of image inputs \cite{b35}. For small datasets, the U-Net structure is expected to have a better performance than traditional convolution neural networks (CNNs) \cite{b35}. The architecture of the MRI segmentation network consists of a down-sampling path (left hand side in Fig. 2) and up-sampling path (right hand side in Fig. 2). The network takes a 256 × 256-sized MRI image as input and 4 levels of down sampling were used. In the first down sampling level, a convolution on the input image using a 3 × 3 kernel size with stride 1, padding size of 1 and Rectified Linear Unit (ReLU) produces 64 feature maps, or channels. Another convolution using a 3 × 3 kernel size with stride 1, padding size of 1 and ReLU is then are applied next. After that, a 2 × 2 max pooling operation was used and the channel size was doubled. Each encoder level has the same convolution and max pooling parts. 1024 feature channels were involved in the last level. In the up-sampling path, a 2 × 2 up-sampling with nearest-neighbor interpolation was used and the channel size was halved. Subsequently, these channels were concatenated with the channels in the down-sampling part which have the same size as up-sampling channels. Two 3 × 3 convolution with stride 1, padding size of 1 and ReLU activation were used. In the end, binary masks are generated by the network, which represent the segmentation results of MRI. A similar network architecture was designed for MSOT image segmentation (Fig. 2). The difference between the MRI segmentation network and the MSOT segmentation network was mainly on the input images and network parameters including learning rate, epochs, and batch size. These hyperparameters are given below. 1) MRI segmentation: learning rate = 0.00001, batch size = 128, training epochs for convergence = 200; 2) MSOT segmentation: learning rate = 0.00005, batch size = 32, training epochs for convergence = 5000. Besides, the convolution part of each encoder and decoder level is different. The MSOT segmentation network has kernel size = 5 × 5 and padding = 2 for the first convolution and  kernel size = 3 × 3 and padding = 1 for the second convolution.

\begin{figure}[!t]
\centerline{\includegraphics[width=\columnwidth]{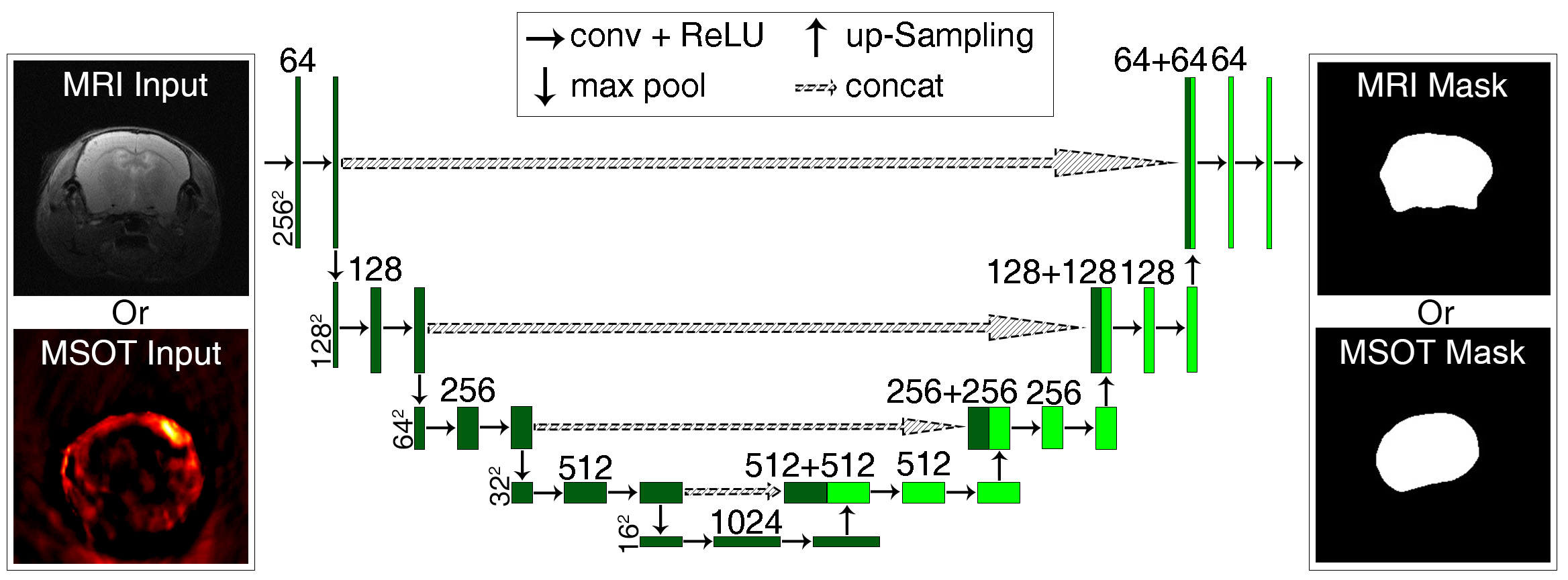}}
\caption{Architecture of MRI and MSOT segmentation networks. A MSOT/MR image sized 256 × 256 pixels (left side) is used as the input of the network. The binary segmentation masks of the brain region (right side) were used as the ground truth. The middle panel represents a U-Net like structure with the boxes denoting the features to be extracted at each step (encoder dark green and decoder light green). The number of channels and the size of features are marked on the top and the left of the box, respectively. The dark green box represents the features from encoder, whereas the light green box represents the features from decoder.}
\label{fig2}
\end{figure}

A binary cross-entropy loss was used as the loss function of both networks. Specifically, for each input MSOT/MR image pixel value $x_{i}$ and each target mask predicting pixel value $y_{i}$, the loss is calculated as

\begin{equation}
\label{eq1}
loss(x,y) = -\frac{1}{N}\sum_{i=1}^{N}[y_{i}logx_{i}+(1-y_{i})log(1-x_{i})],
\end{equation}

where $N$ is the total number of the pixels. For the $i_{th}$ pixel, $x_{i}$ represents the intensity of MSOT/MR image normalized to 1 (1 for brain and 0 for background) and $y_{i}$ represents the index of the $i_{th}$ pixel of MSOT/MR image (1 for brain and 0 for background). The target function, $loss(x,y)$, was minimized using an Adam optimizer \cite{b36}. To keep the image size unchanged, all convolution layers in the network had padding size = 1. 

\subsection{Deep learning-based MSOT-MR image transformation}

After segmentation of the brain images provided by both modalities, the MSOT image was mapped onto the MR image using image transformation (Fig. 3). The target of transformation step was to generate an optimal transformation matrix based on the segmentation results from the previous step. Considering the minimal elastic changes caused by sequential MSOT-MR imaging of the mouse brain, an affine transformation was adopted based on the segmented brain regions. The affine transform is defined as

\begin{equation}
\label{eq2}
J(x,y) = A \cdot I(x,y),
\end{equation}

where the transformation matrix was expressed as the 3 × 3 matrix in \eqref{eq3}:

\begin{equation}
\label{eq3}
A = \begin{bmatrix}
a & b & sx\\ 
c & d & sy\\ 
0 & 0 & 1
\end{bmatrix}.
\end{equation}

More specifically, the parameters \emph{a, b, c, d} indicate the scale, rotation and shear on image, and the parameters \emph{sx, sy} indicate image translation. The positions in the moving space, or projected space, was denoted as $J(x,y) = [x',y',1]^{T}$, whereas the position of the fixed space was denoted as $I(x,y) = [x,y,1]^{T}$. The parameters in the transformation matrix were optimized through training the CNN with a Spatial Transformer Network (STN) \cite{b37}. STN-like architectures have been successfully used in the application such as classification, recognition and image detection \cite{b38,b39}. Typically, a STN consists of three parts: a sampler, a localization net and a grid generator. The localization net yields a regression on transform parameters A and the grid generator generates the transformed images with the same size. The sampler makes the transformation differentiable thus can be used in networks. With these parts, the STN learns a transformation matrix to wrap the input image. In the suggested case, we redesigned the STN to fit the size of the input image (Fig. 3). For STN structure, in the first layer of the STN structures, 20 convolution channels with 5 × 5 convolution kernel, stride 1 and no padding were used with maxpool and ReLU. In the second layer, 20 convolution channels with 5 × 5 convolution kernel, stride 1 and no padding setting with maxpool and ReLU were considered. Finally, these channels were linearized to a 500 × 1 matrix with ReLU and later to the 6 parameters that form the pretransform matrix and generate a pre-transformed MSOT image. This image is then concatenate with the masked MR image and they were trained in a convolution network, which has first a 20 convolution channels level with 5 × 5 convolution kernel, stride 1 and no padding with maxpool and ReLU, then another twenty convolution channels with 5 × 5 convolution kernel, stride 1 and no padding setting with maxpool and ReLU, finally, a linear level to a 500 × 1 matrix with ReLU and later to the 6 parameters that form the final transform matrix. Paired MSOT-MRI datasets from the same mouse brain were used for training the transformation network. Taking the segmented MSOT and MRI images as input, a fully automated registration method was defined.

\begin{figure}[!t]
\centerline{\includegraphics[width=\columnwidth]{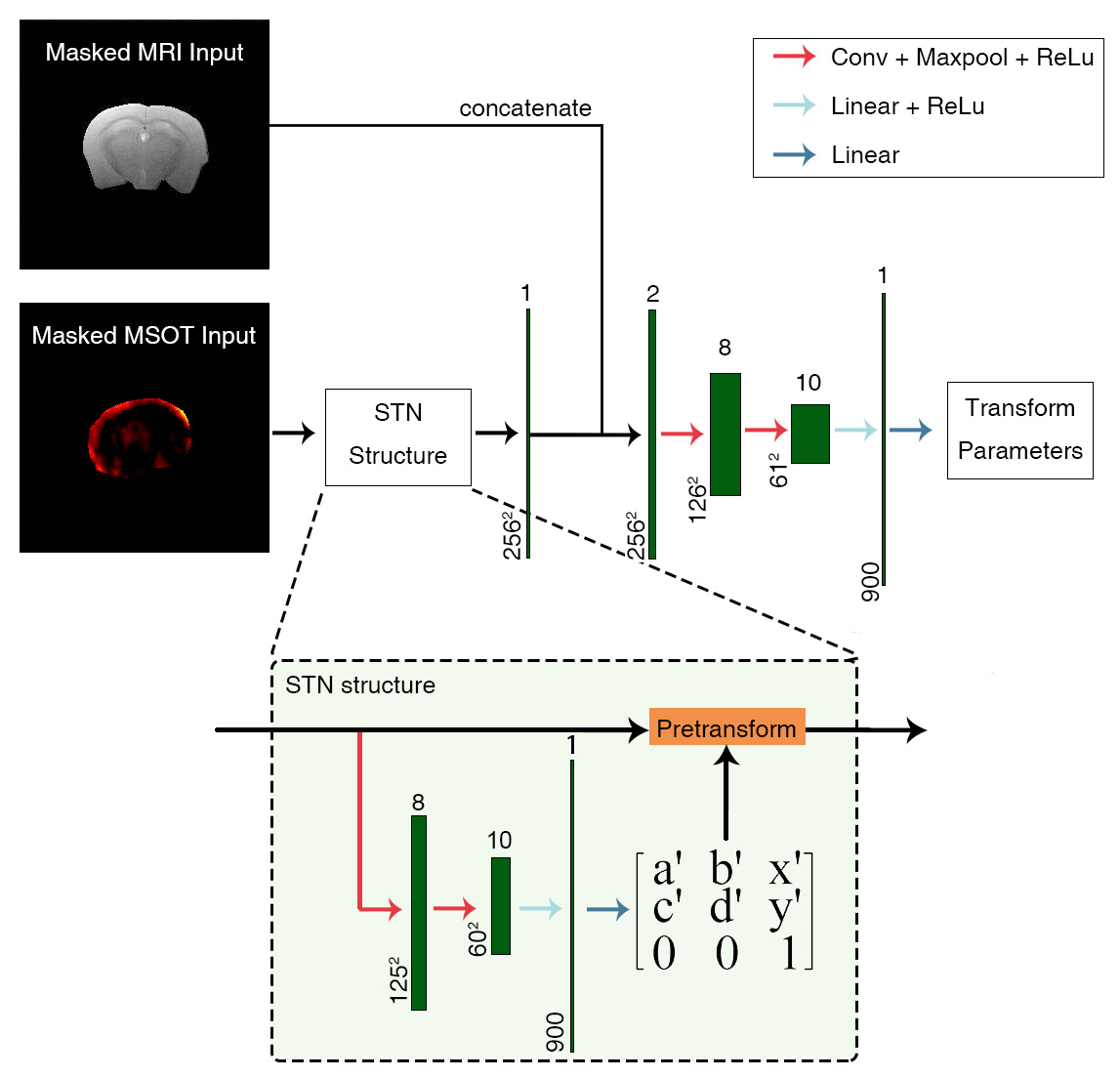}}
\caption{Architecture of the image transformation network. A 256 × 256 masked mouse brain MRI image and a 256 × 256 masked MSOT mouse brain image are used as the input. Firstly, the network performed a spatial transform of the MSOT input to a STN network structure. The STN structure network receives a 256 × 256 image as input and generates a 256 × 256 resampled image as output after training to define the pre-transform parameters. The resampled image was then taken as a new input into a convolution part with MRI input and finally used to predict the parameters in the transformation matrix.}
\label{fig3}
\end{figure}

The loss function of the transformation network has two components. The first one was l1-norm error between the trained transformation parameters and the ground truth parameters which were obtained from a classic MI-based registration method. To improve the final registration accuracy, we added a second component which directly compared the final transformed MSOT and MR images using mean squared error (MSE) as a measure. The complete formulation of the loss function is the following \eqref{eq4}:

\begin{equation}
\label{eq4}
L = SmoothL_{1}(A,A_{g})+\alpha*MSE(x_{MSOT},y_{MSOT}),
\end{equation}

where $SmoothL_{1}$ refers to the Smooth L1-norm error, and $A$ is the generated affine transformation matrix, $A_{g}$ is the ground truth of the transform matrix. The second part of the equation is the MSE loss between $x_{MSOT}$ and $y_{MSOT}$, which stands for the transformed MSOT image and the ground truth respectively. The loss $L$ was minimized using an Adam optimizer \cite{b36}. $\alpha$ denotes a weight parameter that decides which weight the network gives to MSE loss. In practice, $\alpha$ = 0.001 is tested as a default parameter on our dataset and performed good result. The learning rate was set at 0.000001 and the batch size to 128. After 1000 epochs of training, the best parameters corresponding to model with minimum loss on validation set was chosen as the final parameters. We finally use these parameters to train on the whole train set and get the final model.

\subsection{Evaluation}
\subsubsection{Evaluation of MSOT-MRI segmentation}

Firstly, we evaluated the accuracy of the segmentation algorithm on MSOT and MR images. The data used for validation purposes were separate from those used for training the MRI segmentation and MSOT segmentation networks. In total, 415 slices out of all original MSOT dataset were obtained from 5 mice. MSOT images from one mouse were selected as the test set, while another 4 mice were selected as the training and validation sets. The MR images of three mice among 18 mice were selected as the test set and others were selected as training set and validation set.

For both MSOT and MRI segmentation, a Dice coefficient \cite{b40} was introduced to evaluate the obtained result, which is commonly used to assess the accuracy and robustness of image segmentation. The Dice coefficient can be calculated based on the true positive (TP), false positive (FP) and false negative (FN) ratios by comparing pixel by pixel. TP means that the ground truth is positive and the result is positive. FP means the ground truth is negative but the result is positive. FN means the ground truth is negative but the result is positive. Expression of the Dice coefficient is given by \eqref{eq5}:

\begin{equation}
\label{eq5}
Dice = \frac{2TP}{2TP+FP+FN}.
\end{equation}

The value of Dice ranges from 0 to 1, with 0 representing a non-overlapped case and 1 for the highest spatial correlation \cite{b40}. 

\subsubsection{Evaluation of MSOT-MRI registration}
Once image registration is done, it is important to assess the positions of several key anatomical reference landmarks. Thus, an index of Target Registration Error (TRE) \cite{b41} was introduced to quantify the accuracy of MSOT-MRI registration. To measure TRE, several representative reference landmarks were selected on both the original and transformed images. Then, we calculated the distance between each point pairs in two images. The TRE value of a single landmarks is formulated as \eqref{eq6}:

\begin{equation}
\label{eq6}
TRE = \sqrt{(x-x')^{2}+(y-y')^{2}},
\end{equation}

where $x$/$y$ denote as the x/y-axis value of the point on the transformed image and $x'$/$y'$ as the x/y-axis value of the point on the reference image. Regularly, multiple landmarks were selected in which case, the mean value and standard deviation of TRE can be measured. 

\subsubsection{Computational cost}

Regarding the time consumption, during the first step, well-trained MSOT and MRI segmentation networks took about 0.4 second to finish brain segmentation. In the second step, the transformation part took about 0.12 s for each MSOT-MR image pair. In total, the whole registration workflow took less than 1 s which significantly less time compared with the MI-based method or manual registration. The whole training and computation was performed on desktop computer with the CPU i9 - 10980XE (3 GHz) and GPU RTX 3090 (10496 cores).

\section{Results}
\subsection{Segmentation of MSOT and MR images}
The MSOT and MR images were set as the input of the MSOT and MRI segmentation networks, respectively, without any preprocessing (Fig. 4a, e). The network successfully segmented a brain region from the whole head anatomy and background noise in the MSOT or MR image. Such region was represented by a binary mask (Fig. 4c, g). For a comparison, the ground truth of segmentation is illustrated in Figs. 4b and 4f. The brain region in MRI was clearly defined and segmented due to the high soft-tissue contrast (Fig. 4d). Regardless of the appearance of strong background noise in MSOT images, the MSOT segmentation network was capable of outlining the boundary of the brain region (Fig. 4h). The Dice coefficient between MRI mask and its ground truth was 0.991. To be noted, for MSOT segmentation a high Dice value of 0.989 was obtained which further demonstrates the accuracy of the MSOT segmentation networks.

\begin{figure}[!t]
\centerline{\includegraphics[width=\columnwidth]{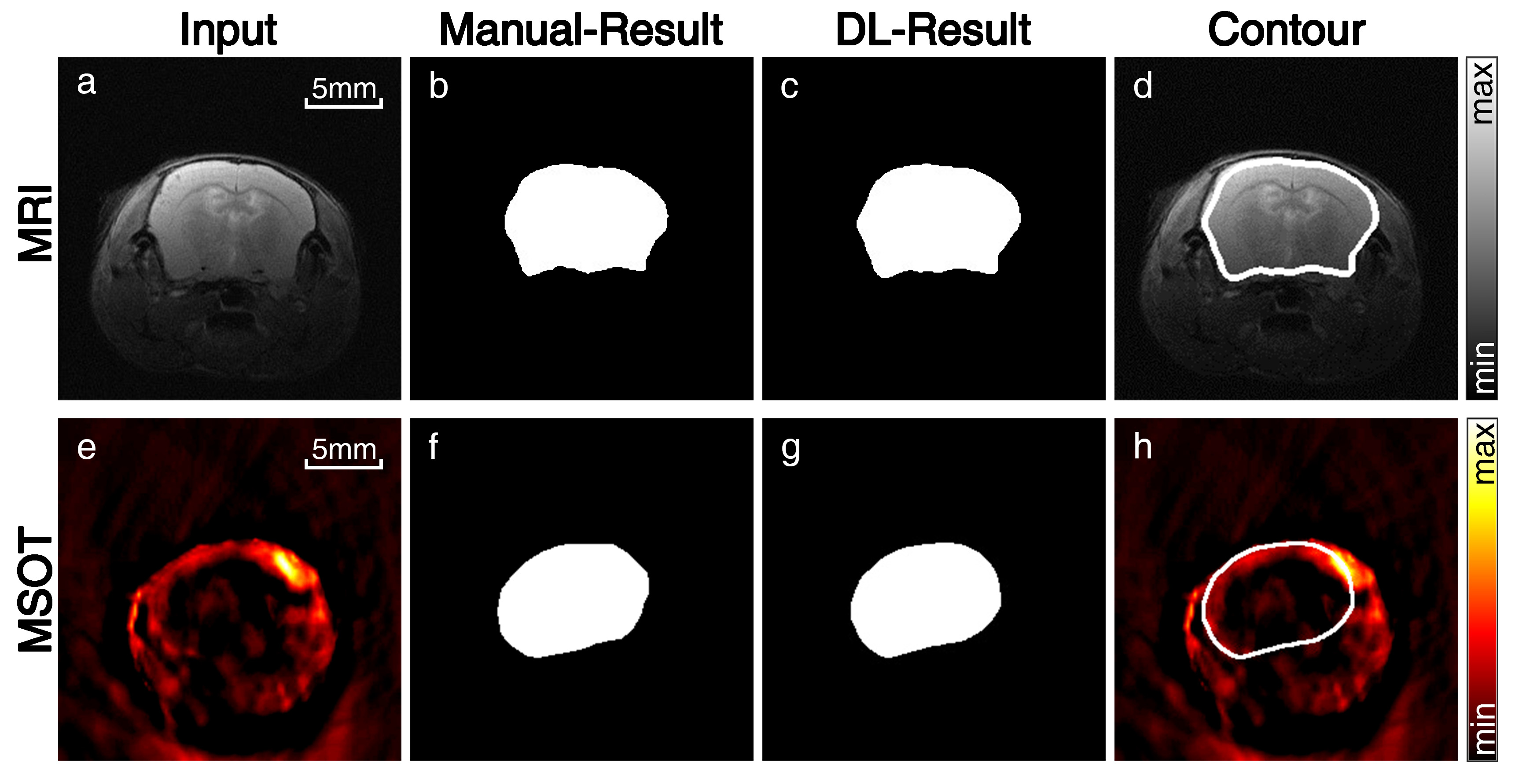}}
\caption{Segmentation results for MSOT and MR images. A raw T$_{2}$-weighted MRI image (a) or a MSOT image (e) were set as the input of the segmentation network structure (both in a coronal view). For a comparison, the ground truth of segmentation was given for each modality (b, f). (c) and (g) show the segmentation results in a form of binary images. (d) and (h) are the overlaid MR and MSOT images with contours illustrating the brain region.}
\label{fig4}
\end{figure}

In order to prevent over-fitting and get full use of the limited dataset for more robust networks, 10-folds cross validation was applied for choosing the best parameters. The Dice coefficients of each fold test are given in Table 1. The means and standard deviations of the 10-fold validation were 0.990 ± 0.001 and 0.979 ± 0.002 for MRI and MSOT, respectively. Both of the mean values were close to 1 which demonstrated the robustness of the MRI segmentation and MSOT segmentation networks.

\begin{table}[ht]
\caption{Cross validation results for MRI and MSOT segmentation network}
\centering
\label{table1}
\begin{tabular}{|c|c|c|}
\hline
Fold number        & \begin{tabular}[c]{@{}c@{}}Dice coefficients for \\ MRI segmentation\end{tabular} & \begin{tabular}[c]{@{}c@{}}Dice coefficients for\\  MSOT segmentation\end{tabular} \\ \hline
1                  & 0.991                                                                            & 0.982                                                                            \\ \hline
2                  & 0.990                                                                            & 0.982                                                                            \\ \hline
3                  & 0.990                                                                            & 0.977                                                                            \\ \hline
4                  & 0.990                                                                            & 0.979                                                                            \\ \hline
5                  & 0.990                                                                            & 0.979                                                                            \\ \hline
6                  & 0.990                                                                            & 0.981                                                                            \\ \hline
7                  & 0.989                                                                            & 0.978                                                                            \\ \hline
8                  & 0.989                                                                            & 0.976                                                                            \\ \hline
9                  & 0.990                                                                            & 0.981                                                                            \\ \hline
10                 & 0.989                                                                            & 0.981                                                                            \\ \hline
Mean               & 0.990                                                                            & 0.979                                                                            \\ \hline
Standard deviation & 0.001                                                                            & 0.002                                                                            \\ \hline
\end{tabular}
\end{table}

\subsection{MSOT-MRI Transformation}
After MSOT and MR images were segmented, the binary masks were used as the input of the follow-up transformation network, which can predict the transformation matrix for mapping the MSOT image onto the MR image. The parameters of the transformation matrix were learned through the transformation network. Then, MSOT images were wrapped by using the generated transformation matrix. Here we compared the final registration results by using the suggested Deep learning (DL)-based method and MI-based method \cite{b23}. For the same coronal-view MSOT slice, the two methods achieved similar results (Fig. 5). The boundaries of both the head and brain regions match well with those of MR images as shown in the overlaid MSOT-MR image pairs (Fig. 5e, f). 

We applied the TRE index to evaluate the final registration results and to quantitatively assess the performance of the suggested DL-based method. Five reference landmarks were manually selected by an expert. As shown in Fig. 5, landmarks of points 1-3 were located at the upper layer of the cerebral cortex, whereas points 4, 5 correspond to the piriform cortex of the brain. These landmarks were selected because of the appearance of vessels at all five points, making them highly visible in MSOT images and recognized in MR images. The TRE values of each landmark point demonstrate the differences between automated method’s results and manual results on that reference point. A smaller TRE values represent a better registration result. The TREs of these points were calculated for the performance evaluation and compared with the MI method (Table 2).

\begin{figure}[!t]
\centerline{\includegraphics[width=\columnwidth]{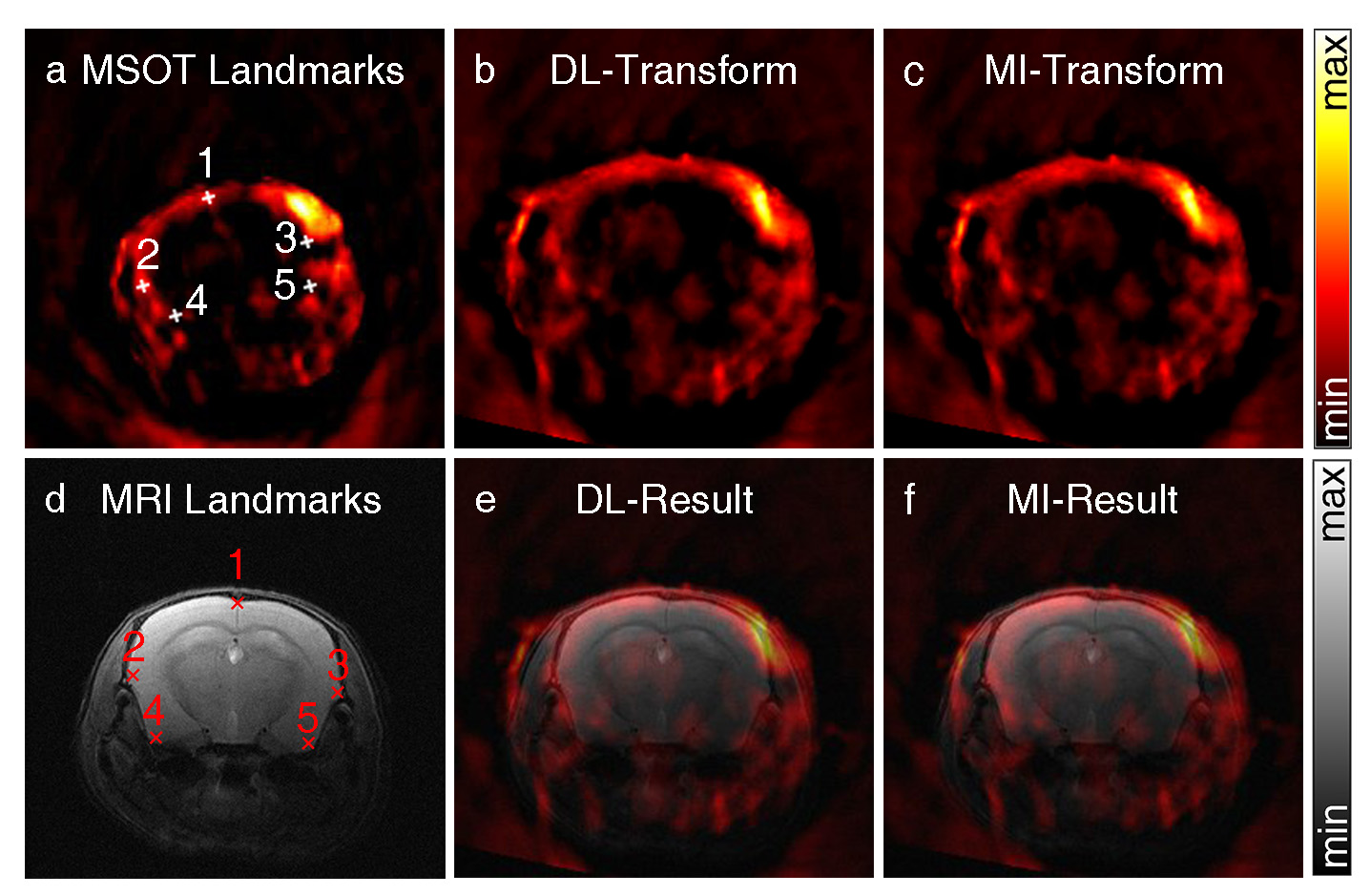}}
\caption{Registered MSOT-MR image pairs. Five MSOT landmarks were marked on the raw reconstructed MSOT image of mouse brain in a coronal view (a) as 1, 2, 3, 4 and 5. The transformed MSOT images using the DL-based and MI-based methods reach a good agreement (b, c). This is further demonstrated in the overlaid MSOT-MR image pairs (e, f), Landmarks 1, 2, 3 are located in the upper layers of the cerebral cortex and landmarks 4, 5 are located at piriform cortex of brain. These points are visible in MSOT images, thus we chose these five points as the landmarks of the suggested transformation.}
\label{fig5}
\end{figure}

\begin{table}[ht]
\centering
\caption{Target Registration Error of MI-based and DL-based registration method}
\label{table2}
\begin{tabular}{|c|c|c|c|c|c|c|}
\hline
Fold & \begin{tabular}[c]{@{}c@{}}Land-\\ mark 1\end{tabular} & \begin{tabular}[c]{@{}c@{}}Land-\\ mark 2\end{tabular} & \begin{tabular}[c]{@{}c@{}}Land-\\ mark 3\end{tabular} & \begin{tabular}[c]{@{}c@{}}Land-\\ mark 4\end{tabular} & \begin{tabular}[c]{@{}c@{}}Land-\\ mark 5\end{tabular} & \begin{tabular}[c]{@{}c@{}}Mean for\\ each image\end{tabular} \\ \hline
1 & 13.466 & 10.415 & 12.354 & 6.916 & 14.344 & 11.499 \\ \hline
2 & 14.681 & 9.981 & 13.709 & 6.552 & 15.332 & 12.501 \\ \hline
3 & 14.916 & 8.848 & 13.270 & 6.019 & 13.983 & 11.407 \\ \hline
4 & 14.290 & 8.861 & 12.513 & 5.388 & 13.670 & 10.944 \\ \hline
5 & 13.948 & 8.539 & 14.123 & 6.047 & 14.183 & 11.368 \\ \hline
6 & 15.145 & 10.804 & 11.801 & 6.861 & 15.259 & 11.974 \\ \hline
7 & 14.607 & 9.884 & 12.627 & 5.980 & 13.226 & 11.265 \\ \hline
8 & 15.043 & 9.229 & 12.093 & 6.234 & 14.079 & 11.336 \\ \hline
9 & 13.154 & 8.738 & 11.426 & 7.698 & 14.647 & 11.133 \\ \hline
10 & 12.712 & 10.691 & 13.536 & 5.764 & 13.118 & 11.164 \\ \hline
Mean & 14.196 & 9.599 & 12.745 & 6.346 & 14.184 & 11.414 \\ \hline
\begin{tabular}[c]{@{}c@{}}MI-\\ based\end{tabular} & 9.501 & 6.106 & 11.857 & 11.688 & 16.468 & 11.124 \\ \hline
\end{tabular}
\end{table}

\section{Discussion}
Similar to many other molecular imaging methods including PET and single photon emission computed tomography (SPECT), there is a rising trend of multi-modal imaging that combines the emerging MSOT technology with a high-resolution structural imaging modality such as MRI or computed tomography (CT) \cite{b42,b43}. In an analogy to the clinically established PET-CT \cite{b44}, the combination of MSOT and MRI can provide both molecular and anatomical information, which is better than PET-CT as CT has no soft tissue contrast. Multi-modal MSOT-MRI can be implemented in either a sequential mode with successive data acquisition \cite{b19} on each modality or by simultaneously using a truly hybrid imaging device \cite{b45}. In both cases, image registration is required. Although several MSOT-MRI registration methods have been previously reported \cite{b19,b20,b21}, manual selection of reference points or manual segmentation was always needed. In contrast, the registration algorithm proposed in this work is based on a deep learning method involving no human intervention if the network is properly trained, thus enables full automation of the whole process. This fully automated workflow requires only raw MSOT and MR images without any pre-processing step like smoothing or de-noising. The speed of the suggested method was also shown to be significantly higher than the previously reported methods \cite{b19,b20,b21}. In addition, no technical background was needed to perform the registration task, thus users can obtain registered images more rapidly and more directly. The deep-learning-based MSOT-MRI method can thus be used in high throughput data analysis in preclinical research as well as early diagnosis for future clinical application.

Both MSOT and MRI segmentation networks have been shown to provide high accuracy and can be used individually. Deep learning techniques have extensively been used for segmentation \cite{b46}. Also, a few deep-learning-based methods have successfully been used for MSOT image reconstruction \cite{b25}. The suggested network then enriches MSOT processing choices. Combined with the MRI segmentation and transformation networks, it provides fast and expert-level MSOT-MRI registration results. As shown in Fig. 5, the DL-based transformation method achieved similar performance compared with a MI-based method. This was further validated by using TRE measures (Table. 2). The computational time of the proposed method was significantly lower than that of existing MI-based registration and manual registration methods. Besides, once the model was trained, no user experience of for interpreting MSOT or MR images is required. Thereby, the DL-based registration approach leads to a high-throughput data analysis in multi-modal MSOT imaging.

Furthermore, the proposed method can potentially be employed for registering 3D datasets of the whole brain \cite{b21,b47,b48,b49}. One thing to be noted is that the registration workflow can be conveniently adapted to other types of MR or MSOT imaging data. In this paper, T$_{2}$-weighted MRI data were used to train the network. Other types of MR images such as diffusion MRI, MR angiography can be also used as training data \cite{b18}. Similarly, the raw MSOT image input used in this work can be replaced by unmixed MSOT images signaling the appearance of specific molecules, e.g., deoxygenated and oxygenated hemoglobin. In this paper, MSOT images for wavelengths 715, 730, 760, 800, 850 nm could be successfully segmented, which indicates that the network structure can handle different types of MSOT images. Thus, an even more wide range of wavelength can be tried in this method without significant changes of network structures, reducing the time cost for different settings at different wavelength in a traditional way. More generally, other modal imaging frameworks such as MSOT-CT, MSOT-ultrasound, or even PET-MRI \cite{b50} could potentially benefit from the suggested DL-based registration workflow. In addition, by segmenting each brain anatomical region, the registration can be applied to the brain atlas. In this way, MSOT images can be directly transformed on the corresponding images from the atlas, thus helping researchers to index the signal position.

Despite the success of the DL-based registration method, there is still some limitations in this work. Firstly, the validated data is still limited to the two-dimensional cross-sectional MSOT images acquired from the ring-shaped transducer array. The correctness of both segmentation and transformation networks has not yet tested on other types of optoacoustic images, in particular the reconstructed MSOT from a spherical transducer array \cite{b51} or optoacoustic microscopic images \cite{b52}. Increasing the training data size or image types could potentially facilitate a wide application of our proposed method. Secondly, the up-sampling procedure of both MRI segmentation and MSOT segmentation networks influences in the final registration results because of the randomness of up-sampling interpolation part \cite{b53}. Network structure without sampling part can avoid this problem. The input of MSOT can also be improved by using fluence correction or non-negative reconstruction \cite{b23,b54,b55,b56}.

\section{Conclusion}
In this work, we proposed a novel deep learning-based method for registering MSOT and MR images. The suggested method features full automation without the need for user experience and saves tremendous time, thus can be used for a high-throughput MSOT data analysis. The simultaneously rendered molecular and structural information of the MSOT-MRI multimodal imaging method can then be very valuable in biomedical research. 

\bibliographystyle{ieeetr}
\bibliography{ref_thesis} 
\addcontentsline{toc}{section}{References}

\end{document}